# A GAP CLEARING KICKER FOR MAIN INJECTOR*

I. Kourbanis[#], P. Adamson, J. Biggs, B. Brown, D. Capista, C.C. Jensen, G.E. Krafczyk, D.K. Morris, D. Scott, K. Seiya, S.R.Ward, G. Wu, M-J Yang, Fermilab, Batavia, IL 60510, U.S.A.

*Abstract*

Fermilab Main Injector has been operating at high Beam Power levels since 2008 when multi-batch slip stacking became operational [1]. In order to maintain and increase the beam power levels the localized beam loss due to beam left over in the injection kicker gap during slip stacking needs to be addressed. A set of gap clearing kickers that kick any beam left in the injection gap to the beam abort have been built. The kickers were installed in the summer of 2009 and became operational in November of 2010. The kicker performance and its effect on the beam losses will be described.

## MULTI-BATCH SLIP STACKING AND INJECTION LOSSES

For multi-batch slip stacking in Main Injector a total of 10 Booster batches each consisting of 81 53 MHz bunches are slip-stacked together resulting in 5 double intensity batches. After recapture an additional Booster batch is injected increasing the total number of Booster batches to 11 (9 to NuMI target).

In Fig. 1 a simulation of the slip stacking is shown after the first group of 5 Booster batches has been decelerated. Some of the beam that has not been decelerated ends up in the injection kicker gap for the batches 6-11(orange squares). This amount of beam is relatively small (less than 1% of total injected beam) but since it is concentrated in the 104-106 injection region of Main Injector represents a localized loss leading to a tunnel activation and prevents us from further increasing the Main Injector beam power. To address this loss, gap clearing kicker magnet have been built [2]. The gap clearing magnets fire just before the injection kicker kicking any beam left in the injection gap to the Main Injector abort.

## KICKER INSTALLATION

In order to have an operational clearing gap kicker in Main Injector we had not only to install the kicker magnets in the tunnel but to provide penetrations for the high voltage cables and cooling lines and a new building to house the thyratron switches, the pulsing forming lines the cooling skid and the controls. The penetrations and the seven gap clearing kicker magnets were installed during the summer 2009 Accelerator shutdown.

For the kicker penetrations new holes had to be drilled on top of the Main Injector tunnel (Fig. 2).

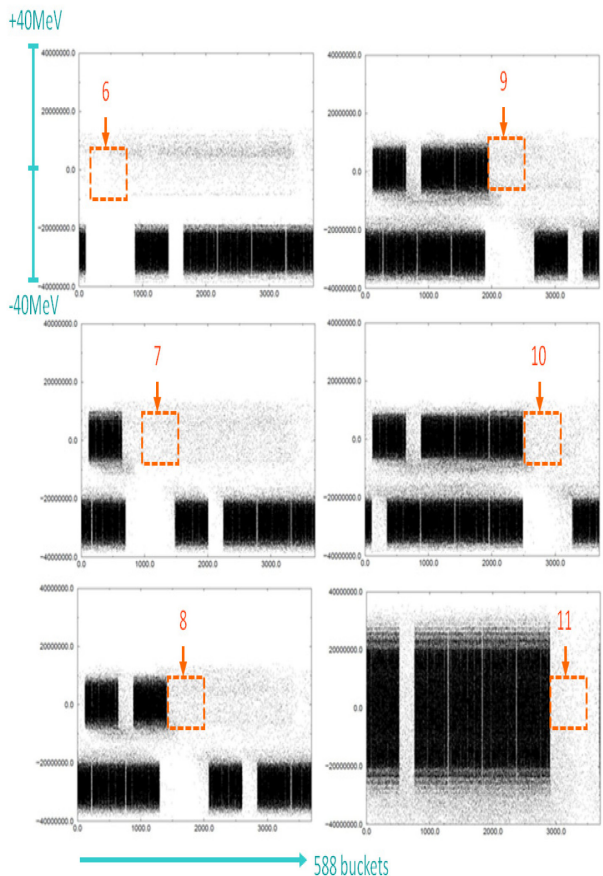

Figure 1: Simulation of multi-batch slip stacking in Main Injector. After 5 Booster batches have been decelerated there some left over beam in the gap where the rest of the 11 batches are injected (red squares).

The kicker building was constructed after the penetrations were installed and the shielding was restored on top of the Main Injector tunnel (Fig. 3). The kicker magnets installed in the Main Injector tunnel are shown in Fig. 4.

The final connections were completed during the Accelerator shutdown of summer 2010 and the controls were finished in October 2010.

___________________________________________
*Work supported by Fermi Research Alliance, LLC under Contrac No. DE-AC02-07CH11359 with the United States Department of Energy.
[#]ioanis@fnal.gov

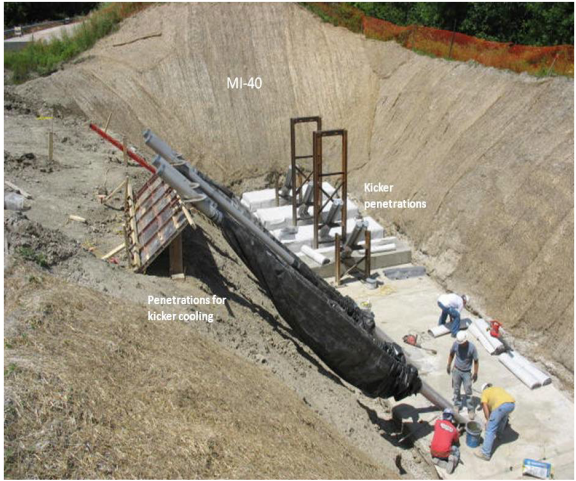

Figure 2: Installation of new kicker penetrations. The shielding has been removed to expose the Main Injector tunnel ceiling. Note gray penetration pipes.

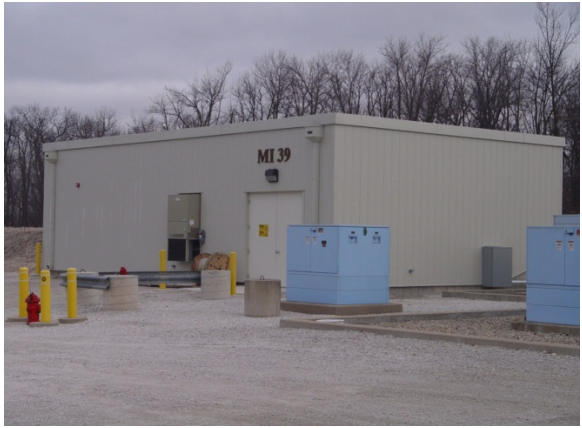

Figure 3: New kicker service building.

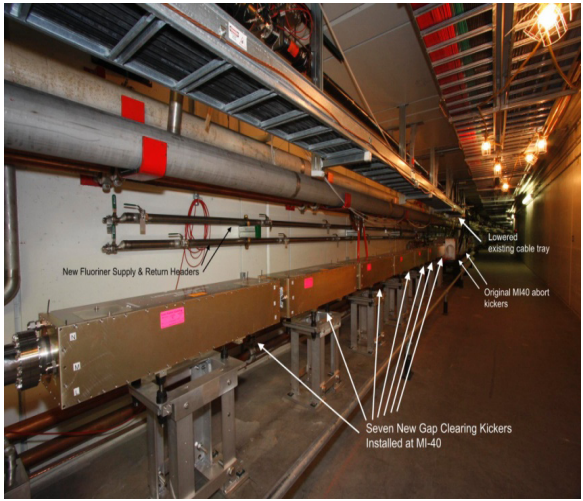

Figure 4: Kicker magnets in the Main Injector tunnel.

## KICKER MEASUREMENTS

Detailed measurements of the Gap Clearing Kicker rise time, fall time and kicker tail were made using the Main Injector bunch by bunch transverse damper [3]. The gap clearing kicker was fired through beam and the damper pickups were used to get bunch by bunch position measurements. The results are shown in Fig. 5,6. Both the rise and fall kicker times are within specifications [2]. The kicker tail will become an issue when we slip stack 12 Booster batches and a "bumper" system to cancel the tail has been designed.

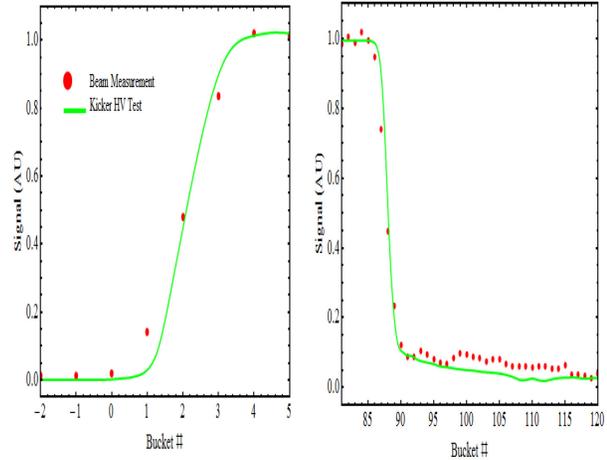

Figure 5: Gap Clearing Kicker rise and fall time measurements. The horizontal scale is in 53 MHz buckets and the vertical scale is in arbitrary units. The red trace is the beam measurements and the green trace is the integrated magnetic field calculated from the kicker voltages. The Kicker Rise and Fall times are within 57 nsec as specified.

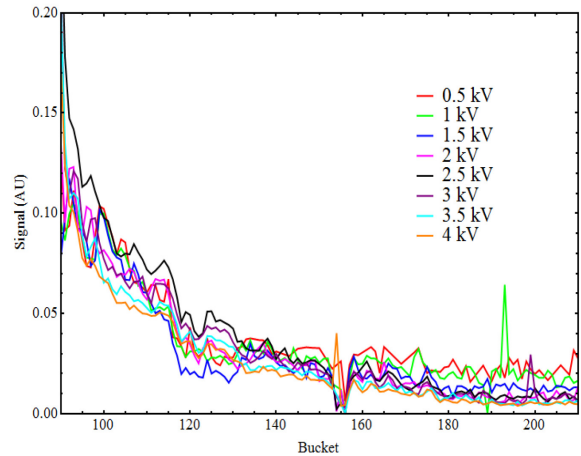

Figure 6: Detailed measurements of the Gap Clearing Kicker tail. The vertical scale is the measured kick in mm and the horizontal scale is in 53 MHz buckets. The different traces correspond to different kicker excitation levels.

## KICKER PERFORMANCE

The Gap Clearing Kicker has been operational since November of 2010. Its effect on Main Injector losses are shown in Fig. 6. The losses at 104,105,106 locations have disappeared. In addition the tunnel activation in the above locations has been reduced by a factor of 6. The Gap Clearing Kicker operation has allowed us to gradually increase the Main Injector beam power already achieving 400 KW (Fig. 8). This was the last project planned to address localized losses in Main Injector due to slip stacking [4].

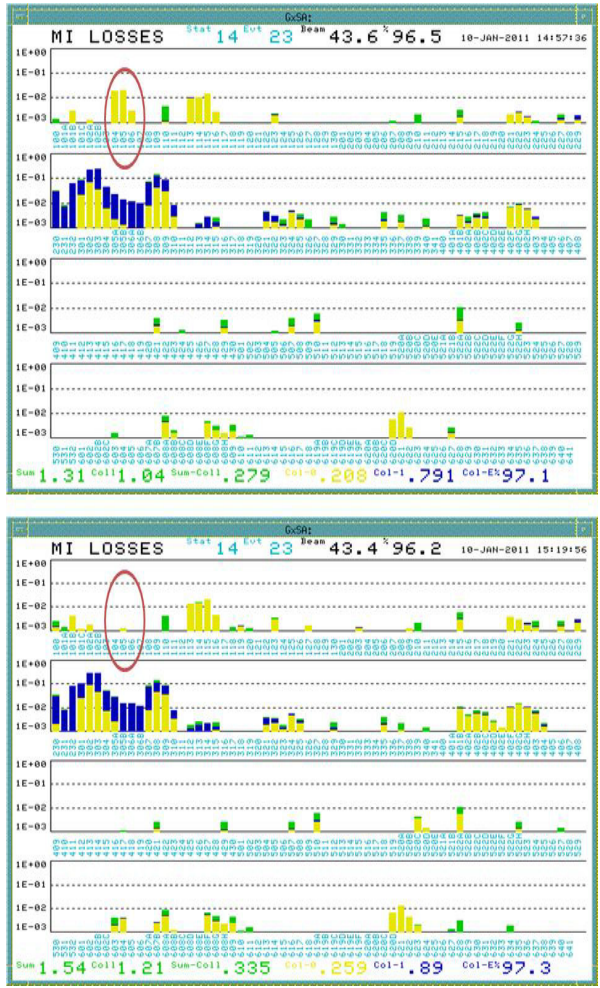

Figure 7: Main Injector loss plot with the Gap Clearing Kicker OFF (top) and ON (bottom). The vertical axis is the integrated loss at each MI beam loss monitor at the end of each cycle (in log scale). The yellow color indicates the contribution to the total loss from injection, the blue from the losses during acceleration and the green from extraction.

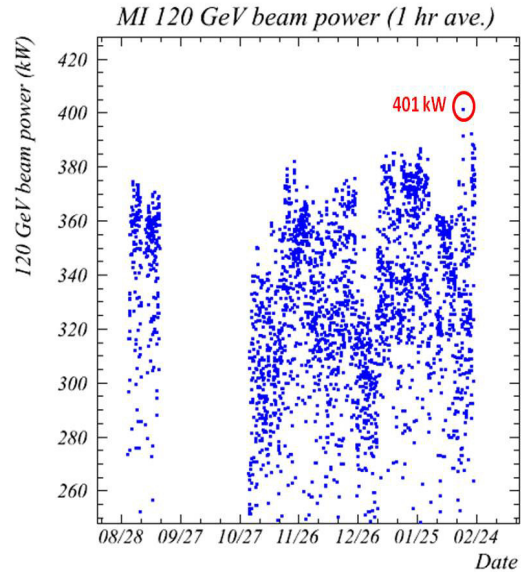

Figure 8: 120 GeV beam power vs. time since August 2010. The gaps in the plots correspond to periods where NuMI neutrino beam was down.

## CONCLUSIONS

A Gap Clearing Kicker is now operational in Main Injector successfully addressing the beam loss from beam left in the injection gap during slip stacking. This has allowed us to increase the beam intensity in Main Injector and to achieve the power goal of 400 KW at 120 GeV.

For the NOvA project the slip stacking will take place in the Recycler so the Gap Clearing Kicker is going to be moved there during the 2012 Accelerator shutdown.